\documentclass[twocolumn,final,prd,superscriptaddress,showpacs,floatfix%
,nofootinbib]{revtex4}

\usepackage{bm}
\usepackage{mathtext}
\usepackage{amssymb, amsmath}
\usepackage{indentfirst}

\usepackage{ifpdf}
\ifpdf
\usepackage{cmap}
\usepackage[pdftex]{color, graphicx}
\usepackage{epstopdf}
\usepackage[pdftex, colorlinks=false, pdfstartview=FitH, hypertexnames=false, unicode,
bookmarks=false]{hyperref}
\else
\usepackage[dvips]{color, graphicx}
\usepackage[hypertex, colorlinks=false, unicode]{hyperref}
\fi

\graphicspath{{Pics/}}

\newcommand{\GF}{G_{\rm F}}
\newcommand{\pF}{p_{\rm F}}
\newcommand{\VF}{V_{\rm F}}
\newcommand{\WF}{\Omega_{\rm F}}

\begin{document}

\title{Radiative decay of keV-mass sterile neutrinos in a strongly magnetized plasma}

\author{Alexandra A.~Dobrynina}
\affiliation{P.\,G.~Demidov Yaroslavl State University,
  Sovietskaya 14, 150000 Yaroslavl,
  Russia}
\affiliation{Max-Planck-Institut f\"ur Physik
  (Werner-Heisenberg-Institut), F\"ohringer Ring 6, 80805 M\"unchen,
  Germany}

\author{Nicolay V.~Mikheev\footnote{Nicolay
Vladimirovich Mikheev passed away on 19 June 2014 before this manuscript
could be completed. The remaining authors hope to have finished this work in
his spirit.}} \affiliation{P.\,G.~Demidov Yaroslavl State University,
  Sovietskaya 14, 150000 Yaroslavl,
  Russia}

\author{Georg G.~Raffelt}
\affiliation{Max-Planck-Institut f\"ur Physik
  (Werner-Heisenberg-Institut), F\"ohringer Ring 6, 80805 M\"unchen,
  Germany}

\date{\today}

\begin{abstract}
The radiative decay of sterile neutrinos with typical masses of 10~keV is
investigated in the presence of a strong magnetic field and degenerate
plasma. Full account is taken of the strongly modified photon dispersion
relation relative to vacuum. The limiting cases of relativistic and
non-relativistic plasma are analyzed. The decay rate in a strongly
magnetized plasma as a function of the electron number density is compared
with the un-magnetized case. We find that a strong magnetic field
suppresses the catalyzing influence of the plasma on the decay rate.
\end{abstract}


\maketitle

\section{Introduction}
\label{sec:intro}

The weak interaction strength of neutrinos as well as their small masses
single them out among all elementary particles. While neutrinos play almost
no role on Earth, their role in astrophysics and cosmology is important and
sometimes dominant. In particular, this pertains to astrophysical cataclysms
like core-collapse supernova explosions or coalescence of neutron stars. In
these phenomena, a dense and hot plasma interacting with a strong neutrino
flux arises. It has become clear that strong magnetic fields of up to
$10^{16}$~Gauss can be generated, exceeding the electron-mass critical field
$B_e = m_e^2 / e \simeq 4.41 \times 10^{13}$~Gauss. Neutrino processes are
also important for the cooling of supernova cores and neutron stars where
neutrinos are emitted from the dense central region. Observations of neutron
stars lead to a wide spread of magnetic-field values, and very large magnetic
fields $B \gtrsim 10^{15}$~Gauss have been identified in some objects called
magnetars~\cite{Duncan:1992hi, Thompson:1995gw}. Therefore, studying
properties and dynamics of such astrophysical phenomena requires a detailed
understanding of quantum processes involving neutrinos under the influence of
a strong magnetic field and relativistic plasma.

The plasma and magnetic field are optically active media and therefore can
significantly influence the photon-neutrino interaction which in vacuum
arises at loop level and turns out to be extremely weak. On the other hand,
the photon-neutrino interaction within a medium can lead to actually observed
effects, notably the neutrino luminosity of a plasma by the $\gamma \to \nu
\bar\nu$ decay~\cite{Adams:1963zzb}. In this process, the plasma has two
effects: it provides photons with an effective mass, enabling the decay
kinematics, and it provides an effective interaction between neutrinos and
photons. On the other hand, the radiative decay of a massive neutrino is
kinematically allowed in vacuum (see, for example, Ref.~\cite{Pal:1981rm} and
references therein). However, an active medium can influence both the decay
amplitude and particle kinematics, and hence, the decay rate can change
significantly~\cite{Raffelt:1996wa, Kuznetsov:2013sea}.

Early studies of the radiative decay of a massless neutrino in a magnetic
field were performed in Refs.~\cite{Galtsov:1972xp, Skobelev:1976at,
Ioannisian:1996pn}. (Note that the process $\nu_i \to \nu_j \gamma$ in the
presence of external fields or media has been called ``radiative decay,''
``Cherenkov effect,'' or ``bremsstrahlung'' in the literature.) The radiative
decay of a massive neutrino $\nu_i \to \nu_j + \gamma$ with $i\neq j$ in the
framework of the Standard Model with lepton mixing was considered in
Ref.~\cite{Gvozdev:1996kx} for electromagnetic fields of different
configurations. In all of these papers, the decay probability was calculated
for low-energy neutrinos ($E_\nu < 2 m_e$) and under the assumption that the
modification of the photon dispersion law can be neglected. In addition, it
was shown that the field-induced amplitude of the ultra-relativistic neutrino
decay in a magnetic field is not suppressed by the smallness of the neutrino
mass, in contrast to vacuum~\cite{Gvozdev:1996kx}.

We recall that with increasing photon energy, its dispersion in a strong $B$
field differs from vacuum and each photon polarization has its own dispersion
law~\cite{Adler:1971wn, Batalin:1971au, Tsai:1974ap}. In particular, the
photon four-momentum~$q^\mu$ can be space like and its square can be
sufficiently large, $|q^2| \gg m_\nu^2$, to allow the transition $\nu_i \to
\nu_j + \gamma$ of a lighter neutrino to a heavier one ($m_i < m_j$). In
other words, the strongly modified photon dispersion law implies that in
practice the radiative decay probability of ultra-relativistic neutrinos in
strong magnetic fields does not depend on the neutrino mass spectrum.

For high-energy neutrinos ($E_\nu \gg m_e$) in a strong constant magnetic
field, the process $\nu \to \nu + \gamma$ was studied in
Ref.~\cite{Gvozdev:1997mc}, taking account of the appropriate photon
dispersion. The same process in a homogeneous magnetic field was considered
in detail in Ref.~\cite{Ioannisian:1996pn} for low-energy neutrinos ($E_\nu <
2 m_e$) and in the kinematical region where the photon dispersion is similar
to vacuum. The neutrino radiative decay was also investigated in
plasma~\cite{Tsytovich:1963,  Oraevsky:1986dt, D'Olivo:1989un, Giunti:1990pp,
Sawyer:1992aj, D'Olivo:1995gy, Hardy:1996gr}. In particular, the decay
probability of a heavier neutrino to a lighter one and a photon in a thermal
medium was calculated in Refs.~\cite{D'Olivo:1989un, Giunti:1990pp} under the
assumption that the particle dispersion relations were not affected by the
plasma.

Later, the study of the neutrino-photon interaction was extended to high
energies in a strongly magnetized electron-positron
plasma~\cite{Chistyakov:1999ii}. In this case, apart from the modified photon
dispersion, large radiative corrections exist near the $e^-e^+$
resonance---otherwise the result is overestimated.

Most recently, the decay of a massive neutrino was analyzed for the
conditions of a strongly magnetized, degenerate electron gas
\cite{Ternov:2013ana}. There are no theoretical restrictions on the existence
of astrophysical objects where both a strong magnetic field and degenerate
plasma can exist. Several objects called magnetars~\cite{Duncan:1992hi,
Thompson:1995gw} have been observed which probably contain such a medium,
i.e., 14 Soft Gamma-Ray Repeaters (SGRs) of which 10 are confirmed and 4 are
candidates as well as 14 Anomalous X-Ray Pulsars (AXPs) with 12 being
confirmed and candidates~\cite{Olausen:2013bpa}. The existence of such
objects motivates the study of elementary processes under extreme conditions.

The main point of our paper is to extend the analysis of
Ref.~\cite{Ternov:2013ana} to include the modified photon dispersion
relation. As a motivation we note that in a strongly magnetized plasma, the
neutrino-photon interaction is mainly determined by electrons occupying the
lowest Landau level. Therefore, the electron chemical potential should
satisfy $\mu_e^2 - m_e^2 < 2 e B$. If the plasma is degenerate ($\mu_e - m_e
\gg T$), the plasma frequency is \cite{Ternov:2013ana, Perez-Rojas:1979fb,
Perez-Rojas:1982}
\begin{equation}
\omega_0^2 =
\frac{2\alpha}{\pi} \, e B \,
\frac{p_{\rm F}}{\sqrt{p_{\rm F}^2 + m_e^2}}\,,
\label{eq:plasmon-freq-squared}
\end{equation}
where $p_{\rm F}$ the electron Fermi momentum. The electron number density in
a strongly magnetized electron gas is $n_e=eB\,p_{\rm F}/(2\pi^2)$
\cite{Landau-Lifshitz-1980-book}. This relation allows us to express the
plasma frequency of Eq.~(\ref{eq:plasmon-freq-squared}) in the form
\begin{equation}
\omega_0 \simeq 37.1~\mbox{keV}
\left ( \frac{n_{30}^2 b^2}{b^2 + 1.3 \, n_{30}^2} \right )^{1/4},
\label{eq:plasmon-freq}
\end{equation}
where $b = B/B_e$ and $n_{30} = n_e / (10^{30} \, \mbox{cm}^{-3})$. Our
benchmark number density $(10^{30}~\mbox{cm}^{-3})$, interpreted here as a
baryon density, corresponds approximately to a mass density of
$10^6~\mbox{g~cm}^{-3}$, where degenerate electrons would still be
nonrelativistic.

For the conditions of interest, a typical scale of $\omega_0$ is therefore
10~keV or larger. Ordinary neutrinos have sub-eV masses so that radiative
decays would not be kinematically possible. Of course, the presence of
electrons implies a weak potential for electron neutrinos of $\sqrt 2 \GF n_e
\simeq 1.27 \times 10^{-7}~\mbox{eV}~n_{30}$ which is a very small effect
compared with the plasma frequency. Therefore, it is the modification of the
photon dispersion relation that tends to be the dominant effect. It is clear
that radiative decays would be of interest only for sterile neutrinos $\nu_s$
with keV masses and above. There has been renewed interest in such particles
recently as a possible warm or cold dark matter candidate
\cite{Boyarsky:2009ix, Raffelt:2011nc, Canetti:2012kh, Drewes:2013gca,
Abazajian:2012ys}. Moreover, the observation of an unexplained 3.5~keV x-ray
line, possibly caused by the $\nu_s \to \nu_a \gamma$ decay of dark-matter
sterile neutrinos, has recently electrified the
community~\cite{Bulbul:2014sua, Boyarsky:2014jta, Boyarsky:2014ska,
Riemer-Sorensen:2014yda, Jeltema:2014qfa}.

Whatever the final verdict on such speculations, we here go through the
exercise of calculating the radiative decay of nonrelativistic sterile
neutrinos in an optically active medium which can be identified with both an
un-magnetized or strongly magnetized plasma. Our main new point beyond the
previous literature is to include the photon dispersion relation
consistently. We limit our discussion to Dirac neutrinos---the Majorana case
should only differ by numerical factors. We neglect the modified active
neutrino dispersion relation in the final state.

We begin in Sec.~\ref{sec:plasma} with the simpler case of an un-magnetized
degenerate plasma for comparison with our main calculation in
Sec.~\ref{sec:magnetic}, the strongly magnetized case. In
Sec.~\ref{sec:conclusions} we summarize our findings.

\section{Un-magnetized plasma}
\label{sec:plasma}

A sterile neutrino $\nu_s$ can mix with an active species and in this way
interact with matter, where~$\theta_s$ is the usual mixing angle. It is
assumed to be very small so that $\nu_s$ essentially coincides with a
propagation eigenstate of mass~$m_s$. For the radiative decay
$\nu_s\to\nu_a\gamma$ in vacuum one finds the probability (or rather decay
rate) \cite{Pal:1981rm}
\begin{equation}
W_{\rm vac} = \frac{9 \alpha \GF^2}{2048\,\pi^4} \,
m_s^5 \sin^2 (2 \theta_s)\,.
\label{eq:vac_propability_2}
\end{equation}
This result pertains to the Dirac case, whereas for Majorana neutrinos the
rate is a factor of~2 larger and then agrees with what is usually stated in
the sterile-neutrino literature~\cite{Boyarsky:2009ix}. We will frequently
use this vacuum result to normalize our results.

Turning next to an unmagnetized electron plasma, the contribution to
radiative decay amplitude is defined by the neutrino-photon interaction via
real electrons. The neutrino-electron interaction is described by the
effective local Lagrangian~\cite{Ioannisian:1996pn}
\begin{equation}
{\cal L}_{\rm eff} = - \frac{\GF}{\sqrt 2} \left [
\bar\Psi_e \gamma^\alpha
\left ( C_V - C_A \gamma_5 \right ) \Psi_e
\right ] j_\alpha\,,
\label{eq:lagrangian_loop}
\end{equation}
where $\Psi_e$ is the electron field. $C_V = \pm 1/2 + 2 \sin^2 \theta_W$ and
$C_A = \pm 1/2$ with the Weinberg angle $\theta_W$ are the vector and
axial-vector coefficients, respectively, which take into account the~$Z$- and
$W$-boson exchange. The plus sign pertains to $\nu_e$, the minus sign to
$\nu_\mu$ and $\nu_\tau$.

\begin{figure}
\centering
\includegraphics[scale=0.8]{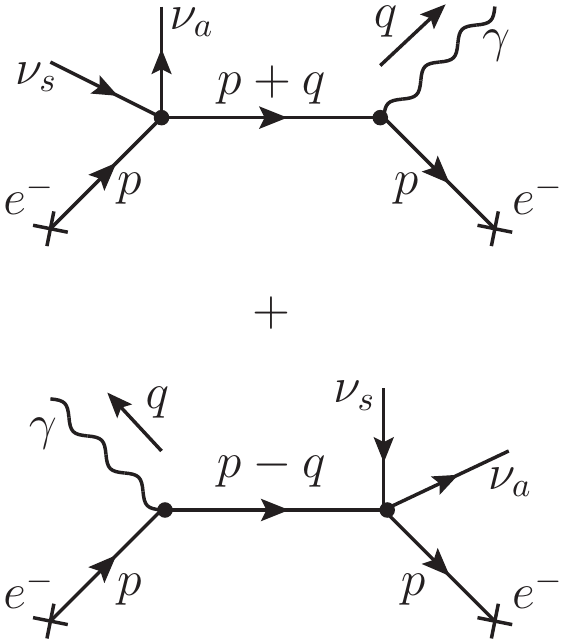}
\caption{
Feynman graphs for the $\nu_s \to \nu_a + \gamma$ decay in plasma.
The crosses attached at the ends of the electron lines signify
that these particles pertain to the plasma. In the magnetized case,
the magnetic field is included on the electron lines.
}
\label{fig:Feynmann}
\end{figure}

\begin{figure}
\centering
\includegraphics[scale=0.8]{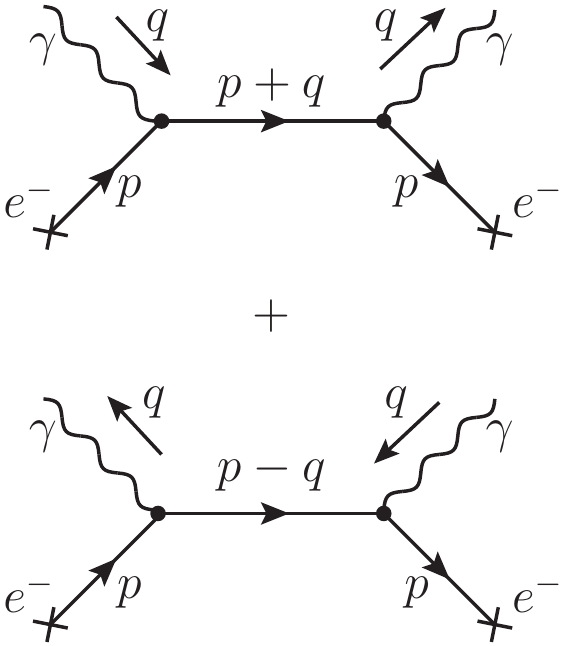}
\caption{
Diagrams for photon forward scattering on plasma electrons
in analogy to Fig.~\ref{fig:Feynmann}.
}
\label{fig:pol}
\end{figure}

The neutrino current $j_\alpha$ in Eq.~(\ref{eq:lagrangian_loop}) describes
the transition of a heavy neutrino $\nu_s$ with a mass of several keV to a
light neutrino $\nu_a$ with a sub-eV mass
\begin{equation}
j_\alpha = \cos \theta_s \sin \theta_s \left [
\bar\nu_a \gamma_\alpha \left ( 1 - \gamma_5 \right ) \nu_s
\right ] .
\label{eq:current_nus}
\end{equation}
The vector current in the Lagrangian~(\ref{eq:lagrangian_loop}) has the same
structure as the standard electron interaction with a photon, ${\cal L}_{\rm
QED} = e \left(\bar\Psi_e \gamma_\alpha \Psi_e \right)A^\alpha$. Therefore,
the decay $\nu_s \to \nu_a + \gamma$ in plasma corresponds to the Feynman
graph shown in Fig.~\ref{fig:Feynmann} which is identical to the one shown in
Fig.~\ref{fig:pol} after one of the photon lines has been replaced by the
neutrino current.


It is well known that the amplitude of the $\gamma \to \gamma$
transition shown in Fig.~\ref{fig:pol} determines the polarization
operator~$\Pi^{\alpha\beta}$ of the photon~\cite{Berestetsky:1982aq,Peskin:1995ev}
\begin{equation}
M_{\gamma \to \gamma} = - \varepsilon^*_\alpha
\Pi^{\alpha\beta} \varepsilon_\beta .
\label{eq:PO-photon-def}
\end{equation}
Therefore, the vector part of the $\nu_s \to \nu_a + \gamma$
amplitude can be expressed in terms of the photon polarization
operator~$\Pi^{\alpha\beta}$ in plasma,
\begin{equation}
M^{(V)}_{\rm pl} = \frac{C_V \,\GF}{e \sqrt 2}
\left ( j_\alpha \Pi^{\alpha\beta} \varepsilon^*_\beta \right ) ,
\label{eq:amplitude-plasma-vector}
\end{equation}
where $\varepsilon_\beta$ is the photon polarization vector.

The corresponding axial-vector contribution is much smaller. In a
non-relativistic plasma one finds explicitly that it is suppressed by a
factor $(C_A/C_V)\,(m_s/m_e)\ll 1$. In a relativistic plasma, $m_e$ is
replaced by the chemical potential $\mu_e$. We conclude that the axial
coupling contributes very little to the process, in analogy to photon
absorption by neutrinos~\cite{Tsytovich:1963} and for plasmon decay into
neutrino pairs~\cite{Kohyama:1986, Braaten:1993jw}.

As mentioned earlier, photons in plasma acquire an effective mass in the form
of the plasma frequency~$\omega_0$. Under a wide range of conditions,
$\omega_0$ is small enough to fulfill the kinematical conditions for
$\nu_s\to\nu_a\gamma$ with $m_s$ of several tens of keV,
\begin{equation}
\omega_0 < m_s \ll m_e .
\label{eq:kinematics}
\end{equation}
We concentrate on a non-relativistic plasma where
\begin{equation}
\omega_0^2 = \frac{4 \pi \alpha n_e}{m_e}\,,
\label{eq:classic}
\end{equation}
where $n_e=\pF^3/(3\pi^2)$ for degenerate electrons. Therefore, the
kinematical condition~(\ref{eq:kinematics}) restricts the Fermi velocity to
$\VF^2<0.25\,(m_s/(10~\hbox{keV}))^{4/3}$. This condition provides an upper
bound $m_s \ll 30$~keV for which the non-relativistic approximation is
appropriate.

Photons in plasma have three polarization modes, one longitudinal
(polarization vector $\varepsilon^\ell$) and two transverse
($\varepsilon^t$). They are the eigenvectors of the polarization
operator~$\Pi_{\alpha\beta}$ and determine the corresponding set of
eigenvalues~$\Pi_\lambda$ ($\lambda = \ell,\, t$). In a non-relativistic
plasma they are $\Pi_t \approx \omega_0^2$ and
$\Pi_\ell\approx\omega_0^2(1-k^2/\omega^2))$,  where $k = |{\bf k}|$ is the
photon momentum.

The probability for $\nu_s\to\nu_a\gamma$ can be written in the form
\begin{eqnarray}\label{eq:propability-pl}
W_{\rm pl}^\lambda&=& \frac{1}{32\pi^2 m_s}
\int Z_{A \lambda} \left | M_{\rm pl}^\lambda \right |^2
\left [ 1 + f_\gamma (\omega) \right ]
\nonumber\\
&& \hspace*{20mm}{}\times
\delta (m_s - k - \omega) \, \frac{d^3 {\bf k}}{k \omega}\,,
\end{eqnarray}
where $f_\gamma (\omega)$ is the photon distribution function. In a cold
plasma ($T \ll \omega_0$), the deviation of the photon stimulation factor $[1
+ f_\gamma (\omega)]$ from unity can be neglected. The factor $Z_{A \lambda}$
accounts for the renormalized wave-function of the photon,
\begin{equation}
Z_{A \lambda}^{-1} = 1 - \frac{\partial \Pi_\lambda}{\partial\omega^2}\,.
\label{eq:RF-photon-renorm}
\end{equation}
The matrix element is largely determined by the vector part of
Eq.~(\ref{eq:amplitude-plasma-vector}). In terms of the eigenvalues and
eigenvectors of the photon polarization operator we find
\begin{equation}
\left | M_{\rm pl}^\lambda \right |^2 =
\frac{\GF^2 \, C_V^2}{16 \pi \alpha} \, \sin^2 (2 \theta_s)
\left [ m_s^2 - q^2 + 4 (p \varepsilon^\lambda)^2 \right ]
\Pi_\lambda^2 .
\label{eq:ME-squared-pl}
\end{equation}
The $\nu_s$ decay probabilities are then found to be
\begin{eqnarray}
W^t_{\rm pl}&=&
\frac{(\GF \,\omega_0^2)^2 \, C_V^2}{128 \pi^2 \alpha}
\sin^2 (2 \theta_s) \, m_s
\left( 1 - \frac{\omega_0^2}{m_s^2} \right)^2\,,
\label{eq:W_t_plasma}\\
W^\ell_{\rm pl}&=&
\frac{(\GF \, m_s^2)^2 \, C_V^2}{64 \pi^2 \alpha} \sin^2 (2\theta_s) \,
\omega_0 \left ( 1 - \frac{\omega_0}{m_s} \right )^2 .
\label{eq:W_l_plasma}
\end{eqnarray}
The rate with a transverse photon coincides with a well-known result in the
limit $\omega_0 \to0$ \cite{Giunti:1990pp}. Besides the different phase
space, the longitudinal case involves a nontrivial wave-function
renormalization factor~$Z_\ell$.

\begin{figure}
\centering
\includegraphics[scale=0.6]{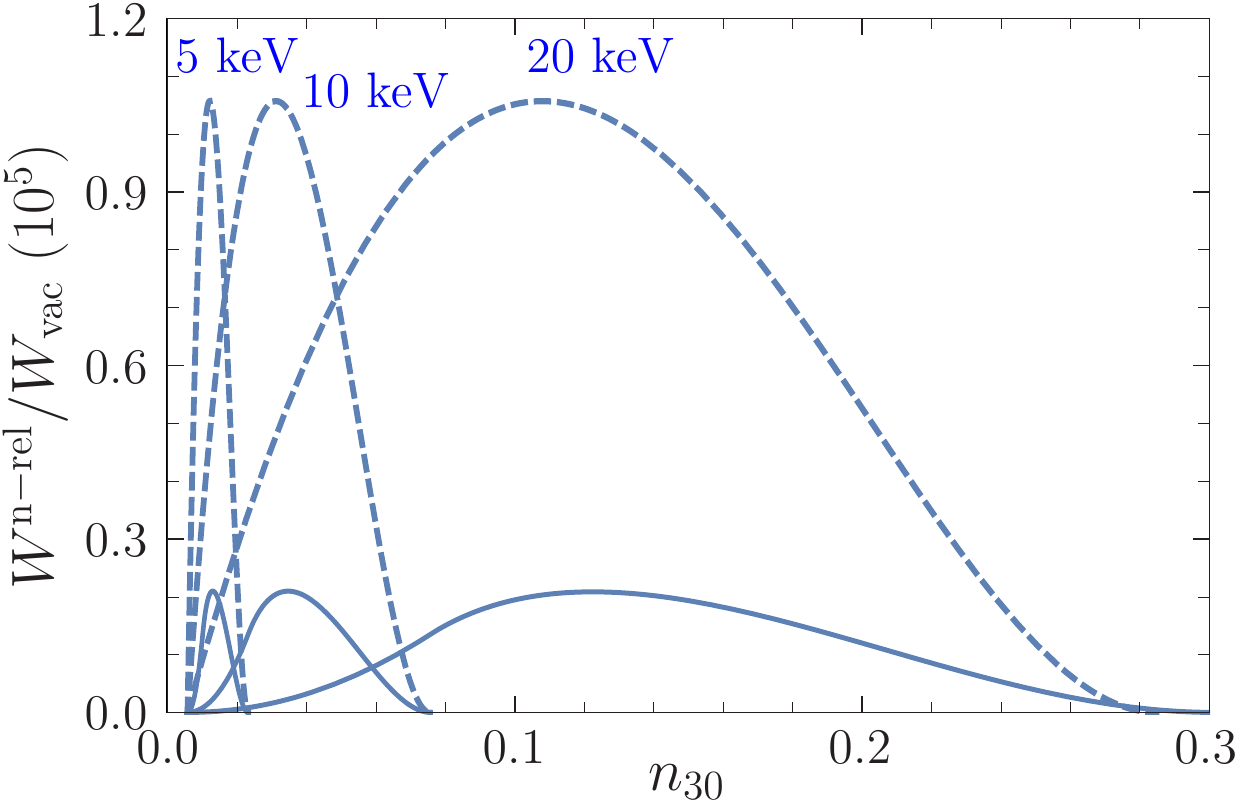}
\caption{Sterile-neutrino radiative decay probability for the indicated mass
values as
a function of the electron density $n_{30} = n_e/(10^{30}~\text{cm}^{-3})$.
{\em Dashed lines:\/} un-magnetized plasma.
{\em Solid lines:\/} strongly magnetized plasma with
$B = B_e = 4.41 \times 10^{13}$~Gauss.}
\label{fig:W_via_n_compare}
\end{figure}

We finally express Eqs.~(\ref{eq:W_t_plasma}) and~(\ref{eq:W_l_plasma}) in
terms of the vacuum rate of Eq.~(\ref{eq:vac_propability_2}) and find
\begin{eqnarray}
&&
W^t_{\rm pl} = W_{\rm vac} \,
\frac{32 \pi^2}{18 \alpha^2} \,
x_0^4 \left ( 1 - x_0^2 \right )^2 ,
\label{eq:W_t_plasma_norm} \\
&&
W^\ell_{\rm pl} = W_{\rm vac} \,
\frac{32 \pi^2}{9 \alpha^2} \,
x_0 \left ( 1 - x_0 \right )^2 ,
\label{eq:W_l_plasma_norm}
\end{eqnarray}
where we have introduced $x_0=\omega_0/m_s$ in terms of the plasma
frequency~(\ref{eq:classic}). The kinematical constraint $x_0<1$ implies that
typically the decay into longitudinal plasmons is much faster than into
transverse ones.

In Fig.~\ref{fig:W_via_n_compare} we show the total rate $W_{\rm pl} = W_{\rm
pl}^\ell + W_{\rm pl}^t$ as a function of the electron density (dashed
lines). The strong catalyzing effect of the plasma is clearly seen with an
enhancement of up to five orders of magnitude compared with vacuum. There is
also a maximum of these functions for an electron density~$n_e$ which moves
to larger number densities with increasing neutrino mass.

\section{Strongly magnetized plasma}
\label{sec:magnetic}

\subsection{Analytic calculation}

In the strongly magnetized case the neutrino-photon interaction is defined by
the same effective Lagrangian~(\ref{eq:lagrangian_loop}) as before. However,
the electron field now is a superposition of solutions of the Dirac equation
in a strong $B$ field. We assume that the hierarchy of plasma parameters is
\begin{equation}
2 eB > \mu_e^2 - m_e^2 \gg T^2
\label{eq:conditions}
\end{equation}
and we take the magnetic field to be oriented along the third axis, i.e.,
${\bf B} = (0,0,B)$.

The neutrino-photon interaction is mainly determined by electrons in the
lowest Landau level~\cite{Skobelev:1975}. Therefore, the electron quantum
field~$\Psi_e$ is an eigenfunction of the projection operator
\cite{Kuznetsov:2013sea,Kuznetsov:2004tb}
\begin{equation}
\Pi_- =
\frac{1 + i (\gamma \varphi \gamma)}{2} =
\frac{1 - i \gamma_1 \gamma_2}{2} ,
\label{eq:projection-operator}
\end{equation}
where $\varphi_{\alpha\beta} = F_{\alpha\beta}/B$ is the dimensionless tensor
of the external magnetic field. We use the short-hand notation $(\gamma
\varphi \gamma) = \gamma^\alpha \varphi_{\alpha\beta} \gamma^\beta$ for the
contraction of Lorentz indices.

The properties of this projection operator reveal an effective equality
\cite{Kuznetsov:2013sea,Kuznetsov:2004tb}
\begin{equation}
\Pi_- \gamma_\alpha \gamma_5 \Pi_- =
(\tilde\varphi \gamma)_\alpha \Pi_- ,
\label{eq:eff-equality}
\end{equation}
where $\tilde\varphi_{\alpha\beta} = \tilde F_{\alpha\beta}/B$ is the dual
dimensionless tensor of the external magnetic field and $(\tilde\varphi
\gamma)_\alpha = \tilde\varphi_{\alpha\beta} \gamma^\beta$. This equality
differs from zero only at $\alpha = 0$ and~3. Therefore, we may transform the
axial-vector electron current in the Lagrangian~(\ref{eq:lagrangian_loop}) to
a vector current of the form
\begin{equation}
\bar\Psi_e \gamma_\alpha \gamma_5 \Psi_e =
\bar\Psi_e \Pi_- \gamma_\alpha \gamma_5 \Pi_- \Psi_e =
\bar\Psi_e (\tilde\varphi \gamma)_\alpha \Psi_e\,,
\label{eq:AV-to-V}
\end{equation}
where $\Pi_- \Psi_e = \Psi_e$ was used. Therefore,
Eq.~(\ref{eq:lagrangian_loop}) becomes
\begin{equation}
{\cal L}_{\rm eff} = e \left (
\bar\Psi_e \, \gamma^\alpha \Psi_e \right ) V_\alpha\,,
\label{eq:newlagrangian}
\end{equation}
where we have introduced the local vector operator
\begin{equation}
V_\alpha = - \frac{\GF}{e \sqrt 2} \left [
C_V \, (\tilde\Lambda j)_\alpha + C_A \, (\tilde\varphi j)_\alpha
\right ]\,.
\label{eq:localoperator}
\end{equation}
The Lorentz tensor $\tilde\Lambda_{\mu\nu} = (\tilde\varphi
\tilde\varphi)_{\mu\nu}$ determines the metric of the two-dimensional
Minkowski subspace of the four-dimensional space-time
\cite{Kuznetsov:2013sea,Kuznetsov:2004tb}.
The direct analogy of the Lagrangian~(\ref{eq:newlagrangian}) with the
electromagnetic case ${\cal L}_{\rm QED} = e \left ( \bar\Psi_e \gamma_\alpha
\Psi_e \right ) A^\alpha$ again allows us to map results from electrodynamics
to neutrino processes.

The diagrams for $\nu_s \to \nu_a + \gamma$ and photon forward scattering in
a strongly magnetized plasma are the same as before (Figs.~\ref{fig:Feynmann}
and~\ref{fig:pol}), except that the electrons are now states in a strong $B$
field. The incoming-photon polarization vector ${\cal E}^{(\lambda)}_\alpha$
($\lambda = 1, 2, 3$) in Fig.~\ref{fig:pol} is replaced with the effective
neutrino current~$V_\alpha$ of Eq.~(\ref{eq:localoperator}) in
Fig.~\ref{fig:Feynmann}. We implicitly assume forward scattering of a photon
of definite polarization~$\lambda$ and the production of a photon with the
same polarization in the sterile-neutrino decay. The basis of photon
polarization vectors ${\cal E}^{(\lambda)}_\alpha$ generally differs from the
basis~$\varepsilon^{(\lambda)}_\alpha$ of the un-magnetized case. The new
polarization operator~$\Pi_{\alpha\beta}$ receives contributions from both
the plasma and the external magnetic field. The eigenvalue problem is now
rather complicated and has not yet been solved in the general
case~\cite{Chistyakov:2012ms}.

However, limiting cases provide simplifications and allow us to find analytic
solutions. A strongly magnetized electron plasma is a beautiful case in
point. In particular, the eigenvalues and eigenvectors of the corresponding
photon polarization operator were found as a power expansion in the inverse
magnetic field strength \cite{Mikheev:2014ila}. In this plasma there are only
two physical states of the photon \cite{Chistyakov:2012ms} which largely
coincide with the photon polarization vectors in the constant uniform
magnetic field~\cite{Kuznetsov:2004tb,Kuznetsov:2013sea}
\begin{equation}
{\cal E}^{(1)}_\alpha \approx
\frac{(q \varphi)_\alpha}{\sqrt{q_\perp^2}}
\qquad\hbox{and}\qquad
{\cal E}^{(2)}_\alpha \approx
\frac{(q \tilde\varphi)_\alpha}{\sqrt{q_\parallel^2}}\,.
\label{eq:PVs-magn-plasma}
\end{equation}
The short-hand $q_\perp^2 = q_\mu \varphi^{\mu\nu} \varphi_{\nu\rho} q^\rho$
and $q_\parallel^2 = q_\mu \tilde\varphi^{\mu\nu} \tilde\varphi_{\nu\rho}
q^\rho$ was used.  The third polarization vector~${\cal E}^{(3)}_\alpha$ is
reduced to the photon four-momentum~$q_\mu$ and can be eliminated by a gauge
transformation~\cite{Chistyakov:2012ms,Shabad:1992vx}.

Therefore, the sterile-neutrino decay amplitude also requires the
corresponding eigenvalues~$\Pi_\lambda$ of the polarization operator with
$\lambda = 1$ and~2 which are \cite{Chistyakov:2012ms,Chistyakov:2008rp}
\begin{eqnarray}
&&
\Pi_1 \approx - \frac{2 \alpha}{\pi} \, \omega \mu_e \, \VF \,
\sqrt{\frac{q^2}{q_\parallel^2}}\,,
\label{eq:eigenvalue1} \\
&&
\Pi_2 \approx \frac{2 \alpha}{\pi} \, e B \, \VF \,
\frac{q^2_\parallel}{\omega^2 - \VF^2 k_3^2}\,.
\label{eq:eigenvalue2}
\end{eqnarray}
Here, $\omega$ is the photon energy, $k_3$ the projection of the photon
momentum on the $B$-field direction, and $\VF = \pF/\mu_e$ the Fermi
velocity. Equations~(\ref{eq:eigenvalue1}) and~(\ref{eq:eigenvalue2}) apply
when the kinematical condition $\omega \lesssim m_s \ll m_e$ is satisfied.

To go further, it is instructive to compare the above eigenvalues under the
plasma conditions of Eq.~(\ref{eq:conditions}). With the values of the
parameters entering Eqs.~(\ref{eq:eigenvalue1}) and~(\ref{eq:eigenvalue2})
close to what is maximally allowed, i.\,e., $\omega \sim m_s$, $k_3 \ll m_s$,
and $q^2,\, q_\parallel^2\sim m_s^2$, one easily obtains
\begin{equation}
\left | \frac{\Pi_1}{\Pi_2} \right | \simeq
\frac{\mu_e m_s}{e B} \lesssim \frac{m_s}{\mu_e} \ll 1\,.
\label{eq:Pi1/Pi2-estimate}
\end{equation}
This means that if both eigenvalues contribute to the decay amplitude with
weights of the same order in~$m_s$, terms with~$\Pi_1$ can be neglected in
the amplitude.

Let us apply the procedure explained above which was successfully worked out
in the case of pure plasma. More precisely, after the replacement of the
photon polarization vector ${\cal E}^{(\lambda)}_\beta \to V_\beta$ by the
neutrino current in Eq.~(\ref{eq:PO-photon-def}), one can express the
sterile-neutrino decay amplitude through the photon polarization
operator~$\Pi_{\alpha\beta}$ as
\begin{equation}
M_{\rm pl+f} = \frac{\GF}{e \sqrt 2} \,
{\cal E}^{(\lambda) *}_\alpha \Pi^{\alpha\beta}
\left [ C_V \, (\tilde\Lambda j )_\beta
+ C_A \, (\tilde\varphi j)_\beta \right ] .
\label{eq:amplitudep_1}
\end{equation}
Comparison of the amplitude~$M_{\rm pl+f}$ obtained with the similar one of
Eq.~(\ref{eq:amplitude-plasma-vector}) calculated in the pure plasma shows
that~$C_A$ appears and can no longer be neglected as will be demonstrated
later. Taking into account the hierarchy of the polarization operator
eigenvalues Eq.~(\ref{eq:Pi1/Pi2-estimate}), mainly photons with the
polarization $\lambda = 2$ are produced in this decay. So, the photon
polarization vector should be identified with~${\cal E}^{(2)}_\alpha$. As a
result, the decay amplitude is
\begin{equation}
M_{\rm pl+f} = \frac{\GF}{e \sqrt 2} \, \Pi_2
\left [ C_V \, ({\cal E}^{(2) *} \tilde\Lambda j)
+ C_A \, ({\cal E}^{(2) *} \tilde\varphi j) \right ]\,,
\label{eq:amplitudep_2}
\end{equation}
where the neutrino current~$j_\alpha$ is given in Eq.~(\ref{eq:current_nus}).
The effective neutrino current~$V_\alpha$ in the strongly magnetized plasma,
where all electrons are in the lowest Landau level, is the projection
of~$j_\alpha$ on the two-dimensional Minkowski subspace and thus is
orthogonal to the other polarization vector with $\lambda = 1$, i.e., $({\cal
E}^{(1)} V) = 0$.

After substituting the polarization vector~${\cal
E}^{(2)}_\alpha$~(\ref{eq:PVs-magn-plasma}) and corresponding
eigenvalue~$\Pi_2$~(\ref{eq:eigenvalue2}) in Eq.~(\ref{eq:amplitudep_2}), we
arrive at the final form of the decay amplitude
\begin{equation}
M_{\rm pl+f} = \frac{\GF \, \Omega_0^2}{e \sqrt 2} \,
\sqrt{q^2_\parallel} \; \frac{
C_V \, (q \tilde\varphi j) + C_A \, (q \tilde\Lambda j)
}{\omega^2 - \VF^2 k_3^2}\,.
\label{eq:amplitudep_3}
\end{equation}
We have introduced the plasma frequency
\begin{equation}
\Omega_0^2 = \frac{2 \alpha eB}{\pi} \, \VF
\label{eq:Omega0-magn-def}
\end{equation}
relevant in the magnetized electron plasma.

The probability of $\nu_s \to \nu_a + \gamma$ requires a phase-space
integration of the amplitude squared~(\ref{eq:amplitudep_3}), including the
appropriate dispersion relations. The magnetized plasma does not strongly
modify the active-neutrino dispersion properties. To get the modified
dispersion relation for a photon with polarization~$\lambda$ one needs to
solve
\begin{equation}
q^2 = \Pi_\lambda\,.
\label{eq:DRs-equation}
\end{equation}
For a photon with $\lambda = 2$ it can be written as
\begin{equation}
\omega^2 = k_3^2 + k_\perp^2 + \Omega_0^2 \,
\frac{\omega^2 - k_3^2}{\omega^2 - \VF^2 k_3^2}\, .
\label{eq:corect_law}
\end{equation}
When the photon momentum vanishes, $k_3^2 = k_\perp^2 = 0$, the photon energy
is $\omega = \Omega_0$ and means the effective photon mass in the magnetized
plasma. Note that the plasma frequency squared~(\ref{eq:Omega0-magn-def})
differs from the similar quantity~(\ref{eq:classic}) defined in the
un-magnetized plasma.

The $\nu_s \to \nu_a + \gamma$ decay can only occur if $\Omega_0< m_s$. This
requirement restricts the Fermi velocity to
\begin{equation}
\VF^2 < 0.01 \left ( \frac{B_e}{B} \right )^2
\left ( \frac{m_s}{10~\hbox{keV}} \right )^4\,.
\label{eq:V_field}
\end{equation}
This expression shows that the radiative decay of a sterile neutrino with
mass 2--20~keV in a highly magnetized plasma requires the latter to be
nonrelativistic.

The decay probability has the standard form of an integral over phase space
of the final-state particles
\begin{eqnarray}
&&
W_{\rm pl+f} = \frac{1}{32\pi^2 m_s}
\int \frac{d^3 {\bf p}_a}{E_a} \, \frac{d^3 {\bf k}}{\omega} \,
\label{eq:propability} \\
&& \hspace*{11mm}
{}\times
\delta^{(4)} ( p_s - p_a - q) \,
\left [ 1 + f_\gamma(\omega) \right ]
Z_{A 2} \left | M_{\rm pl+f} \right |^2\,,
\nonumber
\end{eqnarray}
where $p_s^\mu = (m_s, {\bf 0})$ is the $\nu_s$ four-momentum in its rest
frame, $p_a^\mu = (E_a, {\bf p}_a)$ is the four-momentum of the active
neutrino, and the factor $Z_{A 2}$ defined in Eq.~(\ref{eq:RF-photon-renorm})
accounts for the photon wave-function renormalization.

\begin{figure}[h]
\centering
\includegraphics[scale=0.6]{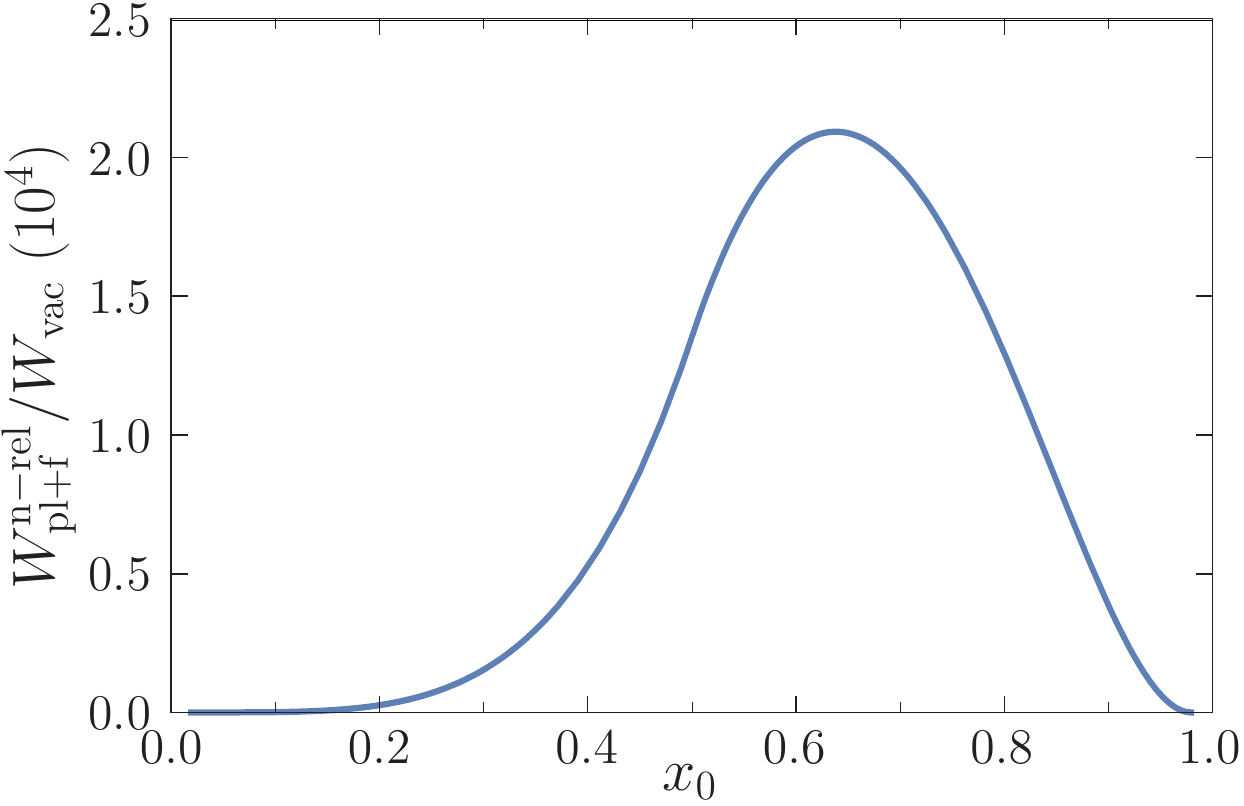}
\caption{Radiative decay probability of sterile neutrinos in a
non-relativistic strongly magnetized plasma
as a function of the reduced plasma frequency $x_0 = \Omega_0/m_s$. }
\label{fig:W_nonrel}
\end{figure}

\begin{widetext}

After performing the integration over the active neutrino momentum~${\bf
p}_a$ and the azimuth angle in the cylindrical momentum frame of the photon,
Eq.~(\ref{eq:propability}) becomes
\begin{equation}\label{eq:propability_cyl}
W_{\rm pl+f} = \frac{1}{32\pi m_s}
\int_{-\infty}^{+\infty} d k_3
\int_0^\infty \frac{d k_\perp^2}{E_a \omega} \,
\delta \, ( m_s - E_a - \omega) \,
\left [ 1 + f_\gamma(\omega) \right ]
Z_{A 2} \left | M_{\rm pl+f} \right |^2\,.
\nonumber
\end{equation}
The remaining integrations are not simple as one should include the
non-trivial photon dispersion relation of Eq.~(\ref{eq:corect_law}) and thus
the $\nu_a$ energy in the form $E_a = m_s - \omega$. It is convenient to
remove the variable~$k_\perp^2$ in favor of~$\omega$ by $dk_\perp^2 = 2
\omega \left |\partial k_\perp^2/\partial\omega^2\right| d\omega$. In the new
variables~$k_3$ and~$\omega$, the integration area is divided into two parts,
leading to
\begin{equation}\label{eq:propability_general}
W_{\rm pl+f} = \frac{1}{16 \pi m_s}
\left(\int_{\Omega_0}^{\WF} d\omega
\int_0^{k_{3{\rm F}}} dk_3 \, F (\omega, k_3)
+\int_{\WF}^\infty d\omega
\int_0^\omega dk_3 \, F (\omega, k_3)\right )\,,
\end{equation}
where $\WF = \Omega_0/\sqrt{1 - \VF^2}$ and $k_{3{\rm F}} = \sqrt{\omega^2 -
\Omega_0^2}/\VF$. The integrand~$F (\omega, k_3)$ in
Eq.~(\ref{eq:propability_general}) can be represented as
\begin{equation}\label{eq:integrand}
F (\omega, k_3) =
\left| \frac{\partial k_\perp^2}{\partial\omega^2} \right |
\frac{\delta\, (m_s - E_a - \omega)}{E_a}
\left [ 1 + f_\gamma (\omega) \right ]
\left ( \left | M_{\rm pl+f} \right |^2 +
\left | M_{\rm pl+f} \right |^2_{k_3 \to -k_3}
\right )\,.
\end{equation}
The calculation of the squared matrix element is not complicated and one
finds
\begin{eqnarray}
\left | M_{\rm pl+f} \right |^2 +
\left | M_{\rm pl+f} \right |^2_{k_3 \to -k_3}
&=& \frac{(\GF \, \Omega_0^2)^2}{8 \alpha\pi} \,
\sin^2 (2\theta_s) \,\frac{q_\parallel^2}{(\omega^2 - \VF^2 k_3^2)^2}
\nonumber\\
&\times&
\left \{
4 m_s^2 \left [ C_A^2 \, \omega^2 + C_V^2 \, k_3^2 \right ]
+q_\parallel^2 \left [
\left ( C_V^2 - C_A^2 \right ) \left ( m_s^2 - q^2 \right ) -
4 C_A^2 \, m_s \omega \right ]
\right \} .
\label{eq:sum_ampl}
\end{eqnarray}
This is our final analytical result for the probability of the
sterile-neutrino radiative decay.

\subsection{Approximations and limiting cases}

In applications it may be more useful to have a simple approximate formula
valid in certain parameter ranges. We adopt $m_s=2$--20~keV as before and $B
=1$--$10\, B_e$ to guarantee strong magnetization. In particular, for $m_s =
10$~keV and $B = 10 \, B_e$ we find
\begin{equation}\label{eq:W_fit}
\frac{W_{\rm pl+f}}{W_{\rm vac}} \approx \frac{\pi^2}{\alpha^2} \,
\left [ 15.93 \, \frac{(1 - x_0)^{0.65}}{x_0^{18.09}}
\exp \left (- 11.79 \, \frac{(1 - x_0)}{x_0} \right )
+ 1168.96 \, (1 - x_0)^{1.46} x_0^{3.88}
\exp \left ( - 0.089 \, \frac{x_0}{1 - x_0} \right )
\right ]\,,
\end{equation}
where $x_0 = \Omega_0 / m_s$. The first function within the square brackets
mainly determines the behavior at large $x_0$ values, while the second one is
for small~$x_0$. The variation of $m_s$ and $B$ in our chosen parameter range
causes only very small changes in the approximation formula. Also, the impact
of the stimulating statistical factor $\left [ 1 + f_\gamma (\omega) \right
]$ is numerically small as in the un-magnetized plasma.

In the same parameter range we can get another approximate representation for
the decay probability. Equation~(\ref{eq:V_field}) reveals that the Fermi
velocity is always small. In the $\VF\ll1$ limit the integrand in
Eq.~(\ref{eq:integrand}) becomes a relatively simple function and can be
integrated analytically,
\begin{equation}\label{eq:W-nonrel-limit}
W_{\rm pl+f}^{\rm n-rel} = W_{\rm vac} \,
\frac{256 \, \pi^2}{25515 \, \alpha^2}
\left ( C_V^2 + C_A^2 \right )
\left [
\theta (2 x_0 - 1) \, F^{\rm n-rel}_1 (x_0) +
\theta (1 - 2 x_0) \, F^{\rm n-rel}_2 (x_0)
\right ]\, ,
\end{equation}
where the functions $F^{\rm n-rel}_{1,2} (x_0)$ are
\begin{eqnarray}
F^{\rm n-rel}_1 (x_0) &=&\frac{2835 x_0^4}{32} \int_0^{1/x_0 -1} dx\,
\left ( 1 - x^2 \right )
\left [ 1 + x_0^2 \left ( 1 - x^2 \right ) \right ]
\left [1 + 3 x^2 - x_0^2 \left (1 - x^2 \right )^2 \right ]
\nonumber \\
&=& - \frac{11}{x_0} + 129 x_0 - 210 x_0^2 + 168 x_0^3
  - 84 x_0^4 - 24 x_0^6 + 32 x_0^8\, ,
\label{eq:F-non-rel-1} \\
F^{\rm n-rel}_2 (x_0) &=&
\frac{2835 x_0^4}{32} \int_0^1 dx\,
\left ( 1 - x^2 \right )
\left [ 1 + x_0^2 \left ( 1 - x^2 \right ) \right ]
\left [1 + 3 x^2 - x_0^2 \left (1 - x^2 \right )^2 \right ]
\nonumber \\
&=& 4 x_0^4 \left ( 21 + 6 x_0^2 - 8 x_0^4 \right ) .
\label{eq:F-non-rel-2}
\end{eqnarray}
\eject
\end{widetext}
The integration variable is $x = \omega/m_s$. The reduced plasma frequency
$x_0 = \Omega_0/m_s$ is restricted to the interval $0 < x_0 < 1$ because of
the decay kinematics. The variation of $W_{\rm pl+f}^{\rm n-rel}$ with $x_0$
is shown in Fig.~\ref{fig:W_nonrel} where both Eqs.~(\ref{eq:W_fit})
and~(\ref{eq:W-nonrel-limit}) coincide numerically.

In Fig.~\ref{fig:W_via_n_compare} we compare the decay rate for the
un-magnetized (dashed lines) and strongly magnetized (solid lines) plasma as
a function of electron density. For the chosen field strength $B=B_e$ the
decay rate is strongly suppressed, but of course it is still much faster than
in vacuum. The maximum decay rate is shifted to somewhat larger electron
densities, reflecting the different dependence of the plasma frequency
on~$n_e$.

At $C_V = C_A = 1$ and $\Omega_0 \ll m_s$ ($x_0 \ll 1$) we reproduce the
result of Ref.~\cite{Ternov:2013ana},
\begin{equation}
W_{\rm pl+f}^{\rm n-rel} =
\frac{256 \, \pi^2}{135 \, \alpha^2} \, x_0^4 \, W_{\rm vac}\,.
\label{eq:W-n-rel-small-x0}
\end{equation}

For relativistic and strongly magnetized conditions, the plasma frequency is
$\Omega_0 \simeq 34.7~\hbox{keV} \, \sqrt{B/B_e}$ and $\nu_s\to\nu_a\gamma$
with $m_s<20$~keV requires $B < B_e/3$. At larger $B$ values the decay mode
$\gamma \to \nu_a + \nu_s$ opens. Therefore, we now implicitly consider
sufficiently heavy sterile neutrinos. In the relativistic limit $\VF \simeq
\sqrt{1 - m_e^2/\mu_e^2} \to 1$ and Eq.~(\ref{eq:propability_general})
simplifies to
\begin{widetext}
\begin{equation}\label{eq:W-pl+f-rel}
W_{\rm pl+f}^{\rm rel} =
\frac{(\GF m_s^2)^2}{64\pi^2\alpha} \,
m_s \, \sin^2 (2\theta_s) \,\left ( C_V^2 + C_A^2 \right )
\frac{x_0^4 \left( 1 + x_0^2 \right )}{1 - e^{- m_s \left ( 1 + x_0^2 \right )/(2 T)}}
\left [ F (x_0, \VF) + \theta \left ( 1 - x_0 \sqrt{\frac{1 + \VF}{1 - \VF}}
\right ) \tilde F (x_0, \VF) \right ]\,.
\end{equation}
Analytical expressions for the functions~$F (x_0, \VF)$ and $\tilde F (x_0,
\VF)$ are give in Appendix~\ref{sec:appendix}. This result further simplifies
in the limiting case of a very small plasma frequency, $x_0 \ll m_e/\mu_e$,
\begin{equation}\label{eq:W-pl+f-rel-small-x0}
W_{\rm pl+f}^{\rm rel} \simeq
\frac{(\GF \Omega_0^2)^2}{64\pi^2\alpha} \,
m_s \, \sin^2 (2\theta_s) \,
\left ( C_V^2 + C_A^2 \right )
\frac{ \ln(2 \mu_e/m_e)-5/4}{1 - e^{- m_s/(2 T)}}\,.
\end{equation}
A simplification also obtains in the opposite limit $x_0 \gg m_e/\mu_e$,
\begin{equation}\label{eq:W-pl+f-rel-large-x0}
W_{\rm pl+f}^{\rm rel} \simeq
\frac{(\GF m_s^2)^2}{64\pi^2\alpha} \,
m_s \, \sin^2 (2\theta_s) \,
\left ( C_V^2 + C_A^2 \right )
\frac{x_0^4}{1 - e^{- m_s \left ( 1 + x_0^2 \right )/(2 T)}}
\left [ \left ( 1 + x_0^2 \right ) \ln \frac{1}{x_0} - \frac{1}{8} \left ( 1
- x_0^2 \right ) \left ( 3 + x_0^2 \right ) \right ]\,.
\end{equation}
Notice that this result applies close to the kinematical limit $m_s$, i.e.,
for $x_0\to 1$.
\vskip18pt

\end{widetext}

\section{Conclusions}
\label{sec:conclusions}

We have studied the radiative decay $\nu_s\to\nu_a+\gamma$ with
cosmologically interesting masses of some 10~keV in a dense magnetized and
un-magnetized electron plasma. Our work goes beyond the previous literature
in that for the first time we have consistently included the modified photon
dispersion relation. The kinematical requirement that the photon effective
mass must be smaller than $m_s$ implies that we should typically restrict the
plasma parameters to non-relativistic conditions.

The decay rate in plasma is much larger than in vacuum because the
neutrino-photon interaction is mediated by plasma electrons instead of
virtual states. In the un-magnetized case, the enhancement is some 5 orders
of magnitude, in detail depending on the electron density. In a strongly
magnetized plasma the enhancement is significantly smaller. A strong $B$
field slows the rate down because the contributing electrons are restricted
to the lowest Landau level. It is also noteworthy that here the electron
axial-current interaction $C_A$ contributes on the same level as the
vector-current $C_V$, in contrast to the un-magnetized case where the vector
current dominates by far. This difference would be especially important if
the final state active flavor is not $\nu_e$ because for $\nu_{\mu}$ and
$\nu_{\tau}$ the vector-coupling constant $C_V$ nearly vanishes.

\section*{Acknowledgments}
We thank I.~S.~Ognev and A.~Ya.\ Parkhomenko for helpful discussions.
Financial support in the framework of the Michail-Lomonosov-Program of the
German Academic Exchange Service (DAAD) and the Ministry of Education and
Science of the Russian Federation (Project No.~11.9164.2014) is acknowledged.
This work was partly supported by the Russian Foundation for Basic Research
(Project No.~14-02-00233-a) and by the German-Russian Interdisciplinary
Science Center (G-RISC) funded by the German Federal Foreign Office via DAAD
under Project No.~P-2013b-19. Partial support by the Deutsche
Forschungsgemeinschaft (DFG) under Grant No.\ EXC-153 (Excellence Cluster
``Universe'') and by the European Union under Grant No.\ PITN-GA-2011-289442
(FP7 Initial Training Network ``Invisibles'') is acknowledged.

\onecolumngrid

\appendix
\section{Probability in the Limit of Relativistic Plasma}
\label{sec:appendix}

The probability of sterile-neutrino radiative decay
in a relativistic magnetized electron plasma
has the form
\begin{eqnarray}\label{eq:W-rel-int}
W_{\rm pl+f}^{\rm rel} &=&
\frac{(\GF \, m_s^2)^2}{64\pi^2\alpha} \,
m_s \, \sin^2 (2 \theta_s)
\left ( C_V^2 + C_A^2 \right )\,
\left [ 1 - e^{- m_s \left ( 1 + x_0^2 \right )/(2 T)} \right ]^{-1}
x_0^4 \left ( 1 + x_0^2 \right )
\nonumber \\
&& \hspace*{22mm}
{}\times \left [
\int_0^a dx \, f (x, x_0)
+ \Theta \! \left ( 1 - x_0 \sqrt{\frac{1 + \VF}{1 - \VF}} \right )
\int_a^1 dx \, f (x, x_0)
\right ]\,,
\end{eqnarray}
where $x = 2 k_3 /[m_s \left ( 1 + x_0^2 \right )]$,
$x_0 = \Omega_0/m_s$, $a = (1 - x_0^2)/(1 + x_0^2)$,
$\Theta (x)$ is the unit-step function,
and the integrand is
\begin{equation}
f (x, x_0) = \frac{1 - x^2}{(1 - \VF^2 x^2)^2} -
\frac{3 + x_0^2}{4} \, \frac{(1 - x^2)^2}{(1 - \VF^2 x^2)^2}\,.
\label{eq:integrand-rel}
\end{equation}
So, there are two simple integrals:
\begin{eqnarray}
F_1 (y, \VF) &=&
\int\limits_0^y \frac{(1 - x^2) \, dx}{(1 - \VF^2 x^2)^2}
= - \frac{y}{2 \VF^2} \left( \frac{1 - \VF^2}{1 - \VF^2 y^2} +
\frac{1 + \VF^2}{2 \VF y} \ln \frac{1 - \VF y}{1 + \VF y} \right) ,
\label{eq:F1-int}\\
F_2 (y, \VF) &=&
\int\limits_0^y \frac{(1 - x^2)^2 \, dx}{(1 - \VF^2 x^2)^2}
= \frac{y}{2 \VF^4}\,\left (
2 + \frac{(1 - \VF^2)^2}{1 - \VF^2 y^2} +
\frac{(3 + \VF^2) \, (1 - \VF^2)}{2 \VF y}
\ln \frac{1 - \VF y}{1 + \VF y} \right )\,.
\label{eq:F2-int}
\end{eqnarray}
The two integrals in Eq.~(\ref{eq:W-rel-int}) are
\begin{eqnarray}
F (x_0, \VF) &\equiv& \int_0^a dx \, f (x, x_0)
= F_{12} (a, \VF) + \frac{1 - x_0^2}{4} \, F_2 (a, \VF)\, ,
\label{eq:F-def} \\
\tilde F (x_0, \VF)&\equiv&\int_a^1 dx \, f (x, x_0)
= F_{12} (1, \VF) - F_{12} (a, \VF)
+ \frac{1 - x_0^2}{4}
\left [ F_2 (1, \VF) - F_2 (a, \VF) \right ] ,
\label{eq:F-tilde-def}
\end{eqnarray}
where it is convenient to use the difference
of the integrals~(\ref{eq:F1-int}) and~(\ref{eq:F2-int}),
\begin{equation}\label{eq:F12-def}
F_{12} (y, \VF) \equiv F_1 (y, \VF) - F_2 (y, \VF)
= - \frac{y}{2 \VF^4} \left [ 2 +
\frac{1 - \VF^2}{1 - \VF^2 y^2} +
\frac{3 - \VF^2}{2 \VF y}
\ln \frac{1 - \VF y}{1 + \VF y} \right ]\,.
\end{equation}
We substitute $x_0^2 =(1 - a)/(1 + a)$
in Eqs.~(\ref{eq:F-def}) and~(\ref{eq:F-tilde-def})
and use the specific values of the functions~(\ref{eq:F2-int})
and~(\ref{eq:F12-def})
\begin{eqnarray}
F_2 (1, \VF) &=&
\frac{1}{2 \VF^4} \left [ 3 - \VF^2
+ \frac{(3 + \VF^2) \, (1 - \VF^2)}{2 \VF}
\ln \frac{1 - \VF}{1 + \VF} \right ]\,,
\\
F_{12} (1, \VF) &=& - \frac{1}{2 \VF^4}
\left [ 3 + \frac{3 - \VF^2}{2 \VF}
\ln \frac{1 - \VF}{1 + \VF} \right ]
\label{eq:F12-unit} \\
\end{eqnarray}
in Eq.~(\ref{eq:F-tilde-def}). We thus arrive at the final
analytical result Eq.~(\ref{eq:W-pl+f-rel}) for the decay
probability of the sterile neutrino.

\eject

\twocolumngrid


\end{document}